\documentstyle[epsfig]{elsart}
\begin{document}
\runauthor{STAR Collaboration }
\begin{frontmatter}
\title{ Energy dependence of $\pi^{\pm}$, $p$ and $\bar{p}$ transverse momentum spectra
      for Au+Au collisions at $\sqrt{s_{\mathrm {NN}}}$~=~62.4 and 200 GeV
}

%%%%% Authorlist taken from STAR Web page %%%%%%
 \author[uic]{B.I.~Abelev},
 \author[pu]{M.M.~Aggarwal},
 \author[vecc]{Z.~Ahammed},
 \author[kent]{B.D.~Anderson},
 \author[dubna]{D.~Arkhipkin},
 \author[jinr]{G.S.~Averichev},
 \author[nikhef]{Y.~Bai},
 \author[indiana]{J.~Balewski},
 \author[uic]{O.~Barannikova},
 \author[uuk]{L.S.~Barnby},
% ...

(STAR Collaboration)

%%%%% Institute list taken from STAR Web page %%%%%%
% ...
\address[uuk]{University of Birmingham, Birmingham, United Kingdom}
% ...
\address[uic]{University of Illinois, Chicago}
% ...
\address[jinr]{Laboratory for High Energy (JINR), Dubna, Russia}
\address[dubna]{Particle Physics Laboratory (JINR), Dubna, Russia}
% ...
\address[indiana]{Indiana University, Bloomington, Indiana 47408}
% ...
\address[kent]{Kent State University, Kent, Ohio 44242}
% ...
\address[nikhef]{NIKHEF and Utrecht University, Amsterdam, The Netherlands}
% ...
\address[pu]{Panjab University, Chandigarh 160014, India}
% ...
\address[vecc]{Variable Energy Cyclotron Centre, Kolkata 700064, India}
% ...

%\author{STAR Collaboration}
%\address{STAR Collaboration}

\date{\today}
\begin{abstract}
We study the energy dependence of the transverse momentum ($p_{\mathrm T}$) 
spectra for charged pions, protons and anti-protons 
for Au+Au collisions  at 
$\sqrt{s_{\mathrm {NN}}}$~=~62.4 and 200~GeV.
 Data are presented at mid-rapidity
($\mid$$y$$\mid$~$<$~0.5) for 0.2~$<$~$p_{\mathrm T}$~$<$~12~GeV/$c$.  
 In the intermediate $p_{\mathrm T}$ region 
(2~$<$~$p_{\mathrm T}$~$<$~6~GeV/$c$), the nuclear modification factor is higher 
at 62.4 GeV than at 200 GeV, while at higher $p_{\mathrm T}$ ($p_{\mathrm T}$~$>$~7~GeV/$c$) 
the modification is similar for both energies.
The $p/\pi^{+}$ and $\bar{p}/\pi^{-}$ ratios for central collisions at 
$\sqrt{s_{\mathrm {NN}}}$~=~62.4 GeV peak at $p_T\simeq2$ GeV/$c$.
In the $p_{\mathrm T}$ range where recombination is expected to dominate, 
the $p/\pi^{+}$ ratios at 62.4 GeV are larger than at 200 GeV, while the $\bar{p}/\pi^{-}$ 
ratios are smaller. For $p_{\mathrm T}$ $>$ 2 GeV/$c$, the $\bar{p}/\pi^{-}$ ratios at the two beam 
energies are independent of  $p_{\mathrm T}$ and centrality indicating that the 
dependence of the $\bar{p}/\pi^{-}$ ratio on $p_{\mathrm T}$ does not change between 62.4 and 200 GeV. 
These findings challenge various models incorporating jet quenching and/or constituent quark coalescence.
\end{abstract}

\end{frontmatter}

\section{Introduction}

%%%%%%%%%%%%%%%%%%%%%%%%%%%%%%%%%%%%%%%%%%%%%%%%%%%%%%%%%%%%%%%%
%What we can learn in Au+Au collisions at different energies
%%%%%%%%%%%%%%%%%%%%%%%%%%%%%%%%%%%%%%%%%%%%%%%%%%%%%%%%%%%%%%%%%%
Experiments at the Relativistic Heavy Ion Collider (RHIC)~\cite{rhicwhitepapers} at
Brookhaven National Laboratory have shown that hadron production at
high  transverse momentum $p_{\mathrm T}$ ($p_{\mathrm T}$~$>$ 6 GeV/$c$) is suppressed for central Au+Au collisions
relative to nucleon-nucleon collisions or peripheral Au+Au
collisions~\cite{star_supression,phenix_pi0}. This suppression is thought to be
related to jet quenching in  dense partonic matter~\cite{jetquenching}. 
At intermediate $p_{\mathrm T}$ (2~$<$~$p_{\mathrm T}$~$<$~6~GeV/$c$), in central
collisions, the baryon to meson ratio is higher than in
peripheral collisions~\cite{baryon_meson_ratio,star_reco}. This feature may be
due to hadronization through the recombination of quarks~\cite{reco}. 

The energy loss by energetic partons traversing the dense medium formed in
high-energy heavy-ion collisions is predicted to be proportional to both the initial
gluon density~\cite{vitev_density} and the lifetime of the 
dense matter~\cite{xnwang_lifetime}. 
The energy
dependence of the nuclear modification factor (NMF, defined later)
significantly constrains parameters in theoretical model calculations. 
The quantitatively large suppression pattern observed at high $p_{\mathrm T}$, 
for both light hadrons and those involving heavy quarks~\cite{starnonphotonic}, 
has renewed interest in the mechanism of energy loss, namely, the relative
contribution of radiative and collisional forms. The dominance of
one over the other depends upon $p_{\mathrm T}$ and
energy~\cite{alam,munshi}. Recently, for a given beam energy a universal
dependence of high $p_{\mathrm T}$ NMF on the number of participating nucleons ($N_{\mathrm part}$) was
proposed as a signature of radiative mechanisms being the dominant energy loss processes ~\cite{vitev_npart}. 
On the other hand, it was suggested that radiative energy loss will break $x_{\mathrm T}$
(=~2~$p_{\mathrm T}$/$\sqrt{s_{\mathrm {NN}}}$) scaling~\cite{scaling}.
Thus, a detailed study of the energy, $p_{\mathrm T}$, and
$N_{\mathrm part}$ dependence of identified hadron production and hadron scaling
properties is needed to continue the investigation of energy loss mechanisms.

In this letter we report the results of such a study performed using identified 
charged pions, protons, and anti-protons for rapidities $\mid$$y$$\mid$~$<$~0.5 and  
$p_{\mathrm T}$~$<$~12~GeV/$c$ for Au+Au at $\sqrt{s_{\mathrm {NN}}}$~=~62.4 
and 200 GeV. The data were taken by the STAR experiment at RHIC ~\cite{starnim}.

Identified particle $p_{\mathrm T}$ spectra at different beam energies will also 
enable the study of the effects of the energy dependence of parton energy loss 
and initial jet production on the produced hadron $p_{\mathrm T}$ spectra. 
At high $p_T$ ($p_T$~${}^>_\sim$~ 6 GeV/$c$), pions are expected to originate dominantly from 
quark jets at $\sqrt{s_{\mathrm {NN}}}$~=~62.4 GeV, while both gluon and quark 
jets contribute substantially to pion production in the same $p_{\mathrm T}$ region at
$\sqrt{s_{\mathrm {NN}}}$~=~200~GeV~\cite{ppdau,ppstrange}. Therefore, a factor of $\sim$ 3
difference in $x_{\mathrm T}$ (for a given $p_{\mathrm T}$) at the two
beam energies may allow the study of the difference in energy loss 
mechanisms for quarks and gluons. This difference in energy loss is due to the non-Abelian
feature of color charge dependence of parton energy loss~\cite{xnwang_nonabelian,th_prob}. 
Alternatively, as $\bar{p}$ production is dominantly from gluon jets,
the $p(\bar{p})$/$\pi$ ratios are  sensitive to quark and gluon jet production in 
heavy-ion collisions~\cite{ko,star_pid200}.
Baryon production relative to meson production is also sensitive to baryon transport and
energy densities. The energy dependence of the baryon-to-meson ratio will 
address the specific prediction of the quark coalescence models of a higher baryon-to-meson 
ratio at $\sqrt{s_{\mathrm {NN}}}$~=~62.4~GeV compared to 200~GeV in the 
intermediate $p_{\mathrm T}$ range~\cite{vitev_62}.

\section{Experiment and Analysis}
%%%%%%%%%%%%%%%%%%%%%%%%%%%%%%%
% General detector description
%%%%%%%%%%%%%%%%%%%%%%%%%%%%%%
The data presented here were taken at RHIC in 2004 using STAR's~\cite{starnim} Time Projection
Chamber (TPC)~\cite{tpc} and a prototype Time-Of-Flight
(TOF)~\cite{startof} detector. The TPC magnetic field was 0.5 Tesla.
%%%%%%%%%%%%%%%%%%%%%%%%%%%%%%%%%%%%%%%%%
%Data type, Analysis details in brief with references
%%%%%%%%%%%%%%%%%%%%%%%%%%%%
%Particle identification
%%%%%%%%%%%%%%%%%%%%%%%%%%%%%
Using a minimally biased trigger (MB), 7.4$\times 10^6$ and 1.4$\times 10^7$
Au+Au events at $\sqrt{s_{\mathrm {NN}}}$~=~62.4 and 200 GeV,
respectively, were analyzed. 1.5$\times 10^7$ 200 GeV Au+Au events from a central trigger were also analyzed, which
corresponds to the top 12\% of the total cross section~\cite{star_pid200}. 
The identified particle spectra for Au+Au collisions at 200 GeV are presented in Ref.~\cite{star_pid200}.
Centrality selection at 62.4 GeV utilized the uncorrected charged particle multiplicity for pseudorapidities 
$\mid\eta\mid$~$<$~0.5, measured 
by the TPC~\cite{star_pid200,pmdftpc}. Ionization energy loss of charged particles in the TPC
was used to identify $\pi^{\pm}$, $p$ and $\bar{p}$ within $\mid\eta\mid$~$<$~0.5 and full azimuth, 
for $p_{\mathrm T}\le 1.1$~GeV/$c$  and $2.5 \le p_{\mathrm T} \le 12$~GeV/$c$. 
Detailed descriptions of TPC particle identification techniques for the low
$p_{\mathrm T}$ range ($0.2~\le p_{T}~\le~2.5$~GeV/$c$) can be found in Ref.~\cite{star_olga}. For
$p_{\mathrm T}~\ge$ 2.5~GeV/$c$, the relativistic rise of ionization
energy loss  was used to identify the $\pi^{\pm}$, $p$ and $\bar{p}$~\cite{star_pid200,pidNIMA}.
The TOF data allowed pion and proton identification up to $p_{\mathrm T}\sim3$~GeV/$c$ for 
$-1\!<\!\eta\!<\!0$  and  $\Delta\Phi$ $\le~\pi/30$ radians~\cite{star_pid200,starcronin}.

%%%%%%%%%%%%%%%%%%%%%%%%%%%
%Corrections
%%%%%%%%%%%%%%%%%%%%%%%%%%%
Identified hadron acceptance and tracking efficiency were studied through Monte Carlo GEANT
simulations~\cite{star_olga,starcronin,antiproton}. At high 
$p_{\mathrm T}$ ($p_{\mathrm T}~\ge$~2.5~GeV/$c$) 
the efficiencies range from 73\% to 87\% and are nearly independent of $p_{\mathrm T}$, but have a 
weak centrality dependence.
Weak-decay feed-down (e.g. $K_{S}^{0}\rightarrow\pi^{+}\pi^{-}$) contributions to the pion
spectra were calculated using  measured $K_{S}^{0}$ and $\Lambda$ yields~\cite{star_reco} 
and a GEANT simulation. The feed-down contributions to the pion spectra were 
found to be $\sim12\%$ at $p_{\mathrm T}=$~0.35~GeV/$c$ and decreasing to $\sim5\%$ for 
$p_{\mathrm T}\ge$ 1~GeV/$c$. The final pion spectra are presented after subtracting 
these contributions. The inclusive $p$ and $\bar{p}$ yields are presented without 
hyperon feed-down corrections to reflect total baryon production.  The corrections
range from $<$ 20\% for p+p and d+Au data~\cite{star_olga,starcronin,antiproton} rising
to $\sim$ 40\% for central Au+Au up to intermediate $p_T$, and are estimated to be
less than 20\% at high $p_T$~\cite{star_pid200}.

%%%%%%%%%%%%%%%%%%%%%%%%%%%%%%%%
%Systematic errors
%%%%%%%%%%%%%%%%%%%%%%%%%%%%%%%%
Systematic errors for the TPC measurements were particle type and $p_{\mathrm T}$ dependent. 
They include: 
uncertainties in efficiency ($\sim$ 8\%); $dE/dx$ position and width (10-20\%);
background from decay feed-down, ghost tracks and PID contamination at high $p_{\mathrm T}$
(8-14\%); momentum distortion due to charge build-up in the TPC volume (0-10\%); 
the distortion of the measured spectra due to momentum resolution (0-5\%).
The systematic errors are added in quadrature. Systematic errors for the TOF data 
for $\pi^{\pm}$, $p$ and $\bar{p}$ spectra are similar at both energies and are 
about  8\%~\cite{starcronin,Lijuan:04}.  The total systematic errors for $\pi^{\pm}$ 
yields at both energies are estimated to be ${}^<_\sim$ 15\%, and those for $p$ and
$\bar{p}$ are ${}^<_\sim$ 25\% over the entire $p_{\mathrm T}$ range studied~\cite{ppdau}.

\section{Transverse momentum distribution}
%%%%%%%%%%%%%% Fig. 1 %%%%%%%%%%%%%%%%%%%%%%%%%%%%%

\begin{figure*}
\begin{center}
\includegraphics[scale=0.6]{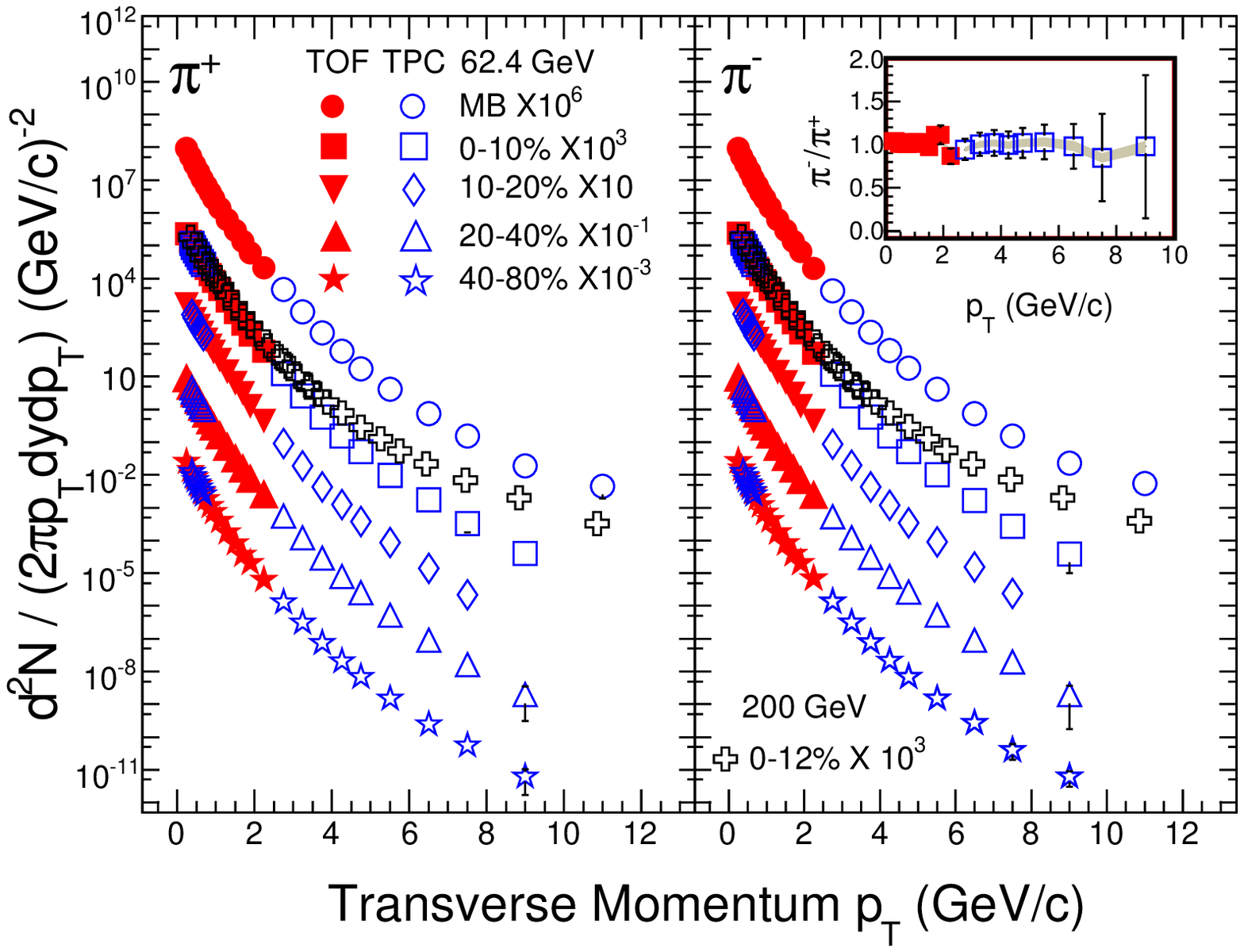}
\includegraphics[scale=0.6]{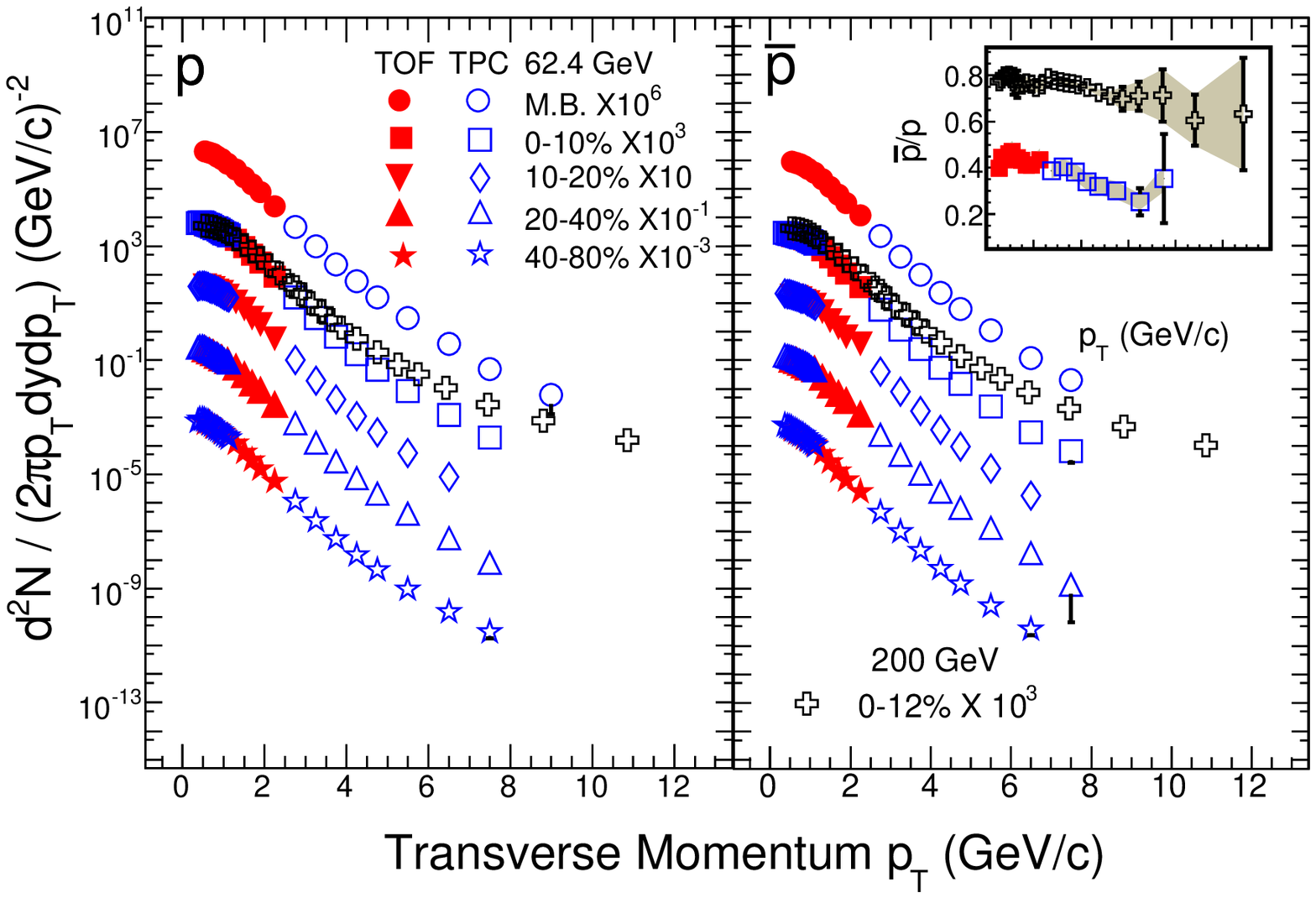}
\caption{Midrapidity ($\mid$$y$$\mid$~$<$~0.5) transverse momentum
spectra for $\pi^{\pm}$, $p$ and $\bar{p}$ for various event
centrality classes for Au+Au  at $\sqrt{s_{\mathrm {NN}}}$~=~62.4~GeV. 
Also shown to study the energy dependence are the
central 0-12\% $\pi^{\pm}$, $p$ and $\bar{p}$ spectra for Au+Au
at $\sqrt{s_{\mathrm {NN}}}$~=~200~GeV. The insets show
$\pi^{-}/\pi^{+}$ at $\sqrt{s_{\mathrm
{NN}}}$~=~62.4 GeV and $\bar{p}/p$ at $\sqrt{s_{\mathrm {NN}}}$ = 62.4 (0-10\%)
and 200 GeV (0-12\%). The errors shown are
statistical, and the shaded bands reflect the systematic errors.}
\label{fig1}
\end{center}
\end{figure*}
%%%%%%%%%% End of Fig.1 %%%%%%%%%%%%%%%%%%%%%%%%%%%%%
Figure~\ref{fig1} shows  $\pi^{\pm}$, $p$ and 
$\bar{p}$ yields for Au+Au at 62.4 GeV for 0.2~$<$~$p_{\mathrm T}$~$<$~12~GeV/$c$ 
and  various collision centralities. 
The hadron
spectra at high $p_{\mathrm T}$ ($p_{\mathrm T}$~$>$~6~GeV/$c$) for 
$\sqrt{s_{\mathrm {NN}}}$~=~62.4~GeV are steeper than the corresponding spectra for 
$\sqrt{s_{\mathrm {NN}}}$~= 200~GeV; comparisons of central collision spectra at both energies
are shown in  Fig.~\ref{fig1}. 
This steepness mostly reflects the difference in initial jet production at the two collision energies.
For high-energy $p$+$p$ and $d$+Au collisions, particle production
at midrapidity is found to follow  $m_{\mathrm T}$ (= $\sqrt{ p_{\mathrm T}^{2} + mass^{2}}$) 
scaling~\cite{ppdau,ppstrange}. Such scaling implies that initial parton  
distributions dominate the particle production process~\cite{cgc}. 
The possibility of $m_{\mathrm T}$-scaling in heavy-ion collisions has been discussed
in Ref.~\cite{cgc}. However, such  $m_{\mathrm T}$-scaling is not
observed in the data for  $\sqrt{s_{\mathrm {NN}}}$~=~62.4~and~200~GeV  Au+Au. 
This will be evident from $p(\bar{p})/\pi$ ratios presented  later.
The absence of $m_{\mathrm T}$-scaling may reflect a
modification of the initial distributions through both partonic and hadronic final state
interactions at RHIC energies.

%%%%%%%%%%% Ratio %%%%%%%%%%%%%
At $\sqrt{s_{\mathrm {NN}}}$ = 62.4 GeV,  $\pi^{-}/\pi^{+}=1.01\pm 0.02$ (stat), independent 
of $p_{\mathrm T}$ within experimental uncertainties (inset of Fig.~\ref{fig1}) and collision centrality 
(not shown). Similar features were observed at 200 GeV~\cite{star_pid200}. 
The  $\bar{p}/p$ ratios  show a slight decrease with $p_{\mathrm T}$ 
(inset of Fig.~\ref{fig1}) and are independent of centrality. The decreasing trend is 
more pronounced at 62.4 GeV~\cite{star_pid200}.
For $p_{\mathrm T}<3$ GeV/$c$, $\bar{p}/p=0.44\pm 0.01$
and $0.77\pm 0.02$ at $\sqrt{s_{\mathrm {NN}}}$ = 62.4 and 200 GeV, respectively. 
For $p_{\mathrm T}>6$ GeV/$c$, $\bar{p}/p=0.29\pm 0.02$ and $0.70\pm 0.05$ 
at $\sqrt{s_{\mathrm {NN}}}$ = 62.4 and 200 GeV, respectively. 

\section{Nuclear modification factor}
The nuclear modification factor is defined relative to
peripheral collisions ($R_{\mathrm {CP}}$) or relative to
nucleon-nucleon collisions ($R_{\mathrm {AA}}$)~\cite{star_supression}: 
$$R_{\rm{CP}}(p_{\rm T})\,=\,\frac{[ d^2N/p_{\rm T}dy dp_{\rm T}/ \langle N_{\rm {bin}}\rangle ]^{central}}{[ d^2N/p_{\rm T}dy dp_{\rm T}/\langle N_{\rm {bin}}\rangle ]^{peripheral}},$$ where
$\langle N_{\mathrm {bin}}\rangle$ is the average number of binary
nucleon-nucleon collisions per event, and
$$R_{\rm{AA}}(p_{\rm
T})\,=\,\frac{d^2N_{\rm{AA}}/dy dp_{\rm T}/\langle N_{\rm {bin}}\rangle}{d^2\sigma_{\rm{pp}}/dy dp_{\rm T}/\sigma_{\rm{pp}}^{\rm {inel}}}.$$
The $\sigma_{\rm{pp}}^{\rm
{inel}}$ are taken to be 36 mb and 42 mb for $\sqrt{s_{\mathrm
{NN}}}$ = 62.4 GeV and 200 GeV, respectively~\cite{pdg}. 
The $d^2\sigma_{\rm{pp}}/dy dp_{\rm T}$ at
200 GeV are from STAR measurements~\cite{ppdau}; for
62.4 GeV we use a parametrization of ISR data~\cite{david} in which the $\pi$ invariant yield for $p$+$p$
at $\sqrt{s_{\mathrm {NN}}}$ = 62.4 GeV is
parameterized as $Ed^3\sigma_{pp\rightarrow \pi X}/d^3p =
A\,(e^{a\cdot p_T^2 + b \cdot p_T}+p_T/p_0)^{-n}$, with 
$A$~= 265.1~mb GeV$^{-2}c^{3}$,$a$ = -0.0129~GeV$^{-2}c^{2}$, $b$~=~0.04975~GeV$^{-1}c$, 
$p_0$~=~2.639~GeV/$c$, and $n$~=~17.95. The
uncertainty in yields associated with this parametrization is
$\sim$ 25\%.

%%%%%%%%%%%%%% Fig. 2 %%%%%%%%%%%%%%%%%%%%%%%%%%%%%
\begin{figure}
\begin{center}
\includegraphics[scale=0.7]{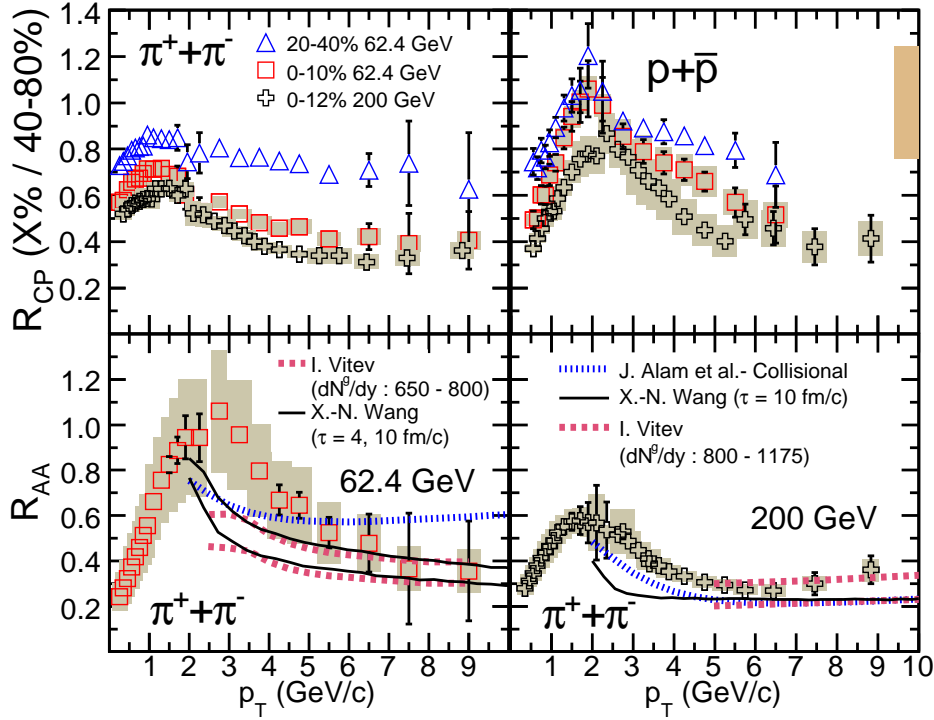}
\caption{Upper panels: Centrality and $p_{\mathrm T}$
dependence of $R_{\mathrm {CP}}$ for $\pi^{+}$+$\pi^{-}$ and
$p$+$\bar{p}$ for Au+Au at $\sqrt{s_{\mathrm {NN}}}$ = 62.4 GeV. For
studying the energy dependence, the corresponding $R_{\mathrm {CP}}$
for central 0-12\% Au+Au at $\sqrt{s_{\mathrm {NN}}}$ = 200 GeV are shown.
Lower panels :
$R_{AA}$ for $\pi^{+}$+$\pi^{-}$  at 62.4 GeV (0-10\%) and 200 GeV (0-12\%)
compared to three model predictions~\cite{vitev_density,xnwang_lifetime,alam} (see text for details).
A 25\% uncertainty is associated with
$d^2\sigma_{\rm{pp}}/dy dp_{\rm T}$
at 62.4 GeV. The error bars are statistical; the shaded bands
are the systematic errors.
The systematic errors for the 20-40\% centrality data
are of similar order as those shown for the 0-10\% data.
The shaded band around  $R_{\mathrm {CP}}$~=~1 at $p_T$ = 10 GeV/$c$ in the top right panel reflects the uncertainty
in $\langle N_{\mathrm {bin}}\rangle$ calculation for 0-10\% collision centrality.
}
\label{fig2}
\end{center}
\end{figure}
%%%%%%%%%% End of Fig.2 %%%%%%%%%%%%%%%%%%%%%%%%%%%%%
Figure~\ref{fig2} (upper panels) shows the $p_{\mathrm T}$, centrality
and $\sqrt{s_{\mathrm {NN}}}$ dependence of $R_{\mathrm {CP}}$ for
$\pi^{+}$+$\pi^{-}$ and $p$+$\bar{p}$ for Au+Au. The
bottom panels show the 
$\pi^{+}$+$\pi^{-}$
$R_{\mathrm {AA}}$ 
for the 0-10\% and 0-12\% centralities at $\sqrt{s_{\mathrm {NN}}}$ = 62.4 and
200~GeV, respectively.
For a given energy
there is a distinct difference in the $p_{\mathrm T}$ dependence between
the $R_{\mathrm {CP}}$ 
for $\pi^{+}$+$\pi^{-}$ and the 
$R_{\mathrm {CP}}$ 
 for $p$+$\bar{p}$
at intermediate $p_{\mathrm T}$. The $R_{\mathrm {CP}}$ for
$p$+$\bar{p}$ has a steeper fall with $p_{\mathrm T}$ compared to
$\pi^{+}$+$\pi^{-}$. At high $p_{\mathrm T}$ the $R_{\mathrm {CP}}$
values are similar for baryons and mesons at both energies. 
The relevance of these measurements for understanding the
energy loss of quarks, gluons and their interaction with the medium
will be discussed together with the $p/\pi^{+}$ and 
$\bar{p}$/$\pi^{-}$ ratios in the next section.
When compared as a function of centrality, a dependence is
observed for $R_{\mathrm {CP}}$ for both $\pi^{+}$+$\pi^{-}$  and 
$p$+$\bar{p}$ at $\sqrt{s_{\mathrm {NN}}}$~=~62.4~GeV. It is found
to be stronger for $\pi^{+}$+$\pi^{-}$. A similar decrease
in $R_{\mathrm {CP}}$ values with increasing collision centrality
was  observed at $\sqrt{s_{\mathrm {NN}}}$~=~200~GeV~\cite{star_pid200}. 
The $R_{\mathrm {CP}}$ values at $\sqrt{s_{\mathrm {NN}}}$~=~62.4~GeV are higher 
than at $\sqrt{s_{\mathrm {NN}}}$~=~200~GeV for
$p_{\mathrm T}$~$<$~7~GeV/c; beyond this $p_{\mathrm T}$ they
approach each other; this  feature may be due to the
interplay of initial jet production and the  gluon density.
For a smaller initial gluon density at the lower energy,  the
$R_{\mathrm {CP}}$ values at the two beam energies may approach each other at high $p_{\mathrm T}$
due to a steeper initial jet spectrum at 62.4 GeV~\cite{xnwang_nonabelian}.

The charged pion $R_{\mathrm {AA}}$ (left bottom panel of
Fig.~\ref{fig2}) for 3.0 $<$ $p_{\mathrm T}$ $<$
8.0 GeV/$c$ at $\sqrt{s_{\mathrm {NN}}}$ = 62.4 GeV decreases with
$p_{\mathrm T}$ and  approaches $\sim$0.35 at
$p_{\mathrm T}$ =  8 GeV/$c$. In contrast the $R_{AA}$
values at $\sqrt{s_{\mathrm {NN}}}$ = 200 GeV are fairly constant
for $p_{\mathrm T}$~$>$~4.0 GeV/$c$ (bottom right panel).  
The difference in the $p_{\mathrm T}$ dependence of $R_{\mathrm {AA}}$ at the
two beam energies is influenced
by the energy dependence of the following: the shape of the initial jet spectrum,  
the parton energy loss, and the relative contributions of quark and gluon jets.
The steeper fall in $R_{\mathrm {AA}}$ with  $p_{\mathrm T}$ 
at 62.4~GeV may be due to the steeper initial jet spectrum. The constant 
value of $R_{\mathrm {AA}}$ at high $p_{\mathrm T}$ for 200 GeV indicates that the 
effect due to the shape of the jet spectrum seems to be compensated by the parton energy loss.
In addition, as  quarks are expected to lose less energy than gluons in
the medium~\cite{xnwang_nonabelian,th_prob}, a higher contribution of quark 
jets at 62.4 GeV compared to 200 GeV for the same $p_{\mathrm T}$ 
($x_{\mathrm T}^{62.4}/x_{\mathrm T}^{200} \sim 3$)
may also cause a difference in the energy dependence of 
$R_{\mathrm {AA}}$ versus $p_{\mathrm T}$.
The differences in the high-$p_{\mathrm T}$ dependence of $R_{\mathrm AA}$ at the two collision energies 
rules out $x_{\mathrm T}$-scaling for Au+Au~\cite{phenix_scaling,star_scaling} 
in contrast to the observations for $p$+$p$~\cite{ppdau}. 
This is expected, as various additional non-perturbative and perturbative
processes for particle production in heavy-ion collisions have
distinct $p_{\mathrm T}$  and $\sqrt{s_{\mathrm {NN}}}$ dependencies.

In Fig.~\ref{fig2} the charged pion $R_{AA}$ are compared to model predictions at 
both energies to study their dependence on the initial gluon
density, the lifetime of dense matter and the mechanism of energy loss. 
The predictions shown do not agree with the data in the region 
2~$<$~$p_{\mathrm T}$~$<$~4~GeV/$c$, indicating that non-perturbative processes may 
dominate hadron production in this $p_{\mathrm T}$ range. 
The dashed curves are from a set of calculations which are 
sensitive to the choice of initial gluon density~\cite{vitev_density,vitev_raa_62}. 
Comparison at high $p_{\mathrm T}$ shows that the initial gluon densities
($dN^g/dy$)
are about 650-800 
and 800-1175 from these calculations for Au+Au at $\sqrt{s_{\mathrm {NN}}}$ = 62.4 and 200 GeV,
respectively.  The lower dashed curves are for higher gluon density. 
In addition, theoretical studies also suggest that for a given 
initial density, the $R_{AA}$($p_{\mathrm T}$) values are sensitive 
to the lifetime ($\tau$) of dense matter formed in heavy-ion 
collisions~\cite{xnwang_lifetime}.  The solid curves are predictions from
Ref.~\cite{xnwang_lifetime} at  $\sqrt{s_{\mathrm {NN}}}$ = 62.4 and 200 GeV
with $\tau$ = 10 fm/$c$ (i.e. larger than the typical system size of $\sim$ 6-7 fm). 
For 62.4 GeV, also shown is a prediction with $\tau$ = 4 fm/$c$ (upper solid line). 
The comparison at high $p_{\mathrm T}$ shows
that, for this model, the lifetime of the dense matter formed in Au+Au collisions 
at $\sqrt{s_{\mathrm {NN}}}$ = 62.4 and 200 GeV is comparable or larger than the system size.
Further insight to the mechanism of energy loss is obtained by comparing the data to 
theoretical predictions (dotted curves) of $R_{AA}$ from models that consider only collisional energy
loss~\cite{alam}. For $\sqrt{s_{\mathrm {NN}}}$ = 200 GeV,
the model predictions of $R_{AA}$ at high $p_{\mathrm T}$ are close to the measured values and
similar to corresponding $R_{AA}$ values from models based on only a radiative mechanism for 
parton energy loss.
However, collisional energy loss model overpredicts the experimental $R_{AA}$ values 
at $\sqrt{s_{\mathrm {NN}}}$ = 62.4 GeV and shows a stronger dependence on beam energy
compared to models based on the radiative process of parton energy loss.

%%%%%%%%%%%%%% Fig. 3 %%%%%%%%%%%%%%%%%%%%%%%%%%%%%
\begin{figure}
\begin{center}
\includegraphics[scale=0.40]{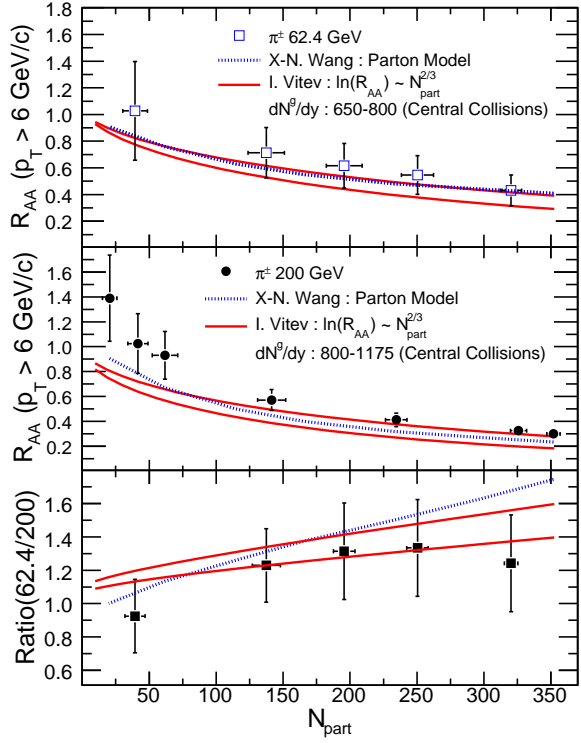}
\caption{$R_{AA}$($p_{\mathrm T}$ $>$ 6 GeV/$c$)
versus $N_{\mathrm part}$ 
for charged pions for Au+Au  at 62.4 GeV
and 200 GeV, and their ratio. The error bars are statistical and systematic
errors added in quadrature. The solid curves are results of calculations 
with radiative energy loss for two different initial gluon
densities in central collisions at both energies and then following the dependence 
of $\ln (R_{\mathrm AA})$ $\propto$ $N_{\mathrm part}^{2/3}$~\cite{vitev_npart}.
The dotted curves are theoretical calculations based on a 
parton model (see text for details)~\cite{xnwang_npart}.}
\label{fig3}
\end{center}
\end{figure}
%%%%%%%%%% End of Fig.3 %%%%%%%%%%%%%%%%%%%%%%%%%%%%%

The centrality dependence of $R_{AA}$ at high $p_{\mathrm T}$ may
provide information on the path length dependence of parton energy loss 
in heavy-ion collisions. Figure~\ref{fig3} shows $R_{AA}$ ($p_{\mathrm T}$ $>$ 6 GeV/$c$) 
as a function of $N_{\mathrm part}$ for $\pi^{+}$+$\pi^{-}$ for 
Au+Au  at 62.4 and 200 GeV. The $R_{AA}$ values 
decrease with $N_{\mathrm part}$ at both energies.
The data are compared to results of two types of model calculations.
The solid curves are from a model that uses a radiative energy loss mechanism
for partons propagating through the medium formed in heavy-ion collisions~\cite{vitev_npart}.
The model assumes the parton production cross section to be a power law type and a parton
energy loss that depends on the initial gluon density, the path length
traveled by the parton, and the transverse area of the region of the
collision. Such a model predicts the centrality dependence of high $p_{\mathrm T}$ 
$R_{\mathrm AA}$ at a given
beam energy to be of the form 
$\ln (R_{\mathrm AA})$ $\sim$ $N_{\mathrm part}^{2/3}$~\cite{vitev_npart}. 
The calculations are done for a set of two different gluon densities at both energies. 
The data follow the predicted dependence at both energies down to low values of $N_{\mathrm part}$.

The dotted curves in Fig.~\ref{fig3} are results from a pQCD based parton model in which 
parton interactions with the medium formed in heavy-ion collisions 
are reflected through the modification of its fragmentation function~\cite{xnwang_npart}. 
The partons in the medium lose energy by induced gluon radiation. 
In such models, the parton energy loss depends upon the 
local gluon density and the total distance of parton propagation~\cite{xnwang_npart}. 
These predictions are  in reasonable agreement with the data for most of 
the centrality classes studied.

The difference between the two models becomes clearer 
when we compare the ratio of  $R_{\mathrm AA}$ ($p_{\mathrm T}$~$>$~6~GeV/$c$) 
values at 62.4 and 200 GeV with the ratios from data. This is shown in
the bottom panel of Fig.~\ref{fig3}. 
For the data, the  $R_{\mathrm AA}$ versus
$N_{\mathrm {part}}$ at 200 GeV is first parametrized by a polynomial function
and then the ratios of $R_{\mathrm AA}$(62.4)/$R_{\mathrm AA}$(200) are calculated.
The pQCD based parton model overpredicts the measurements for the most central collisions. 
For the most peripheral collisions measured, the model calculations from Ref.~\cite{vitev_npart}
slightly overpredict the data. It will be interesting to compare the $R_{\mathrm AA}$ versus
$N_{\mathrm {part}}$ for models with collisional energy loss and see if they provide
further constraint on mechanism of energy loss of partons in heavy-ion collisions.

\section{Baryon-to-meson and anti-baryon-to-baryon ratios}

\begin{figure}
\begin{center}
\includegraphics[scale=0.5]{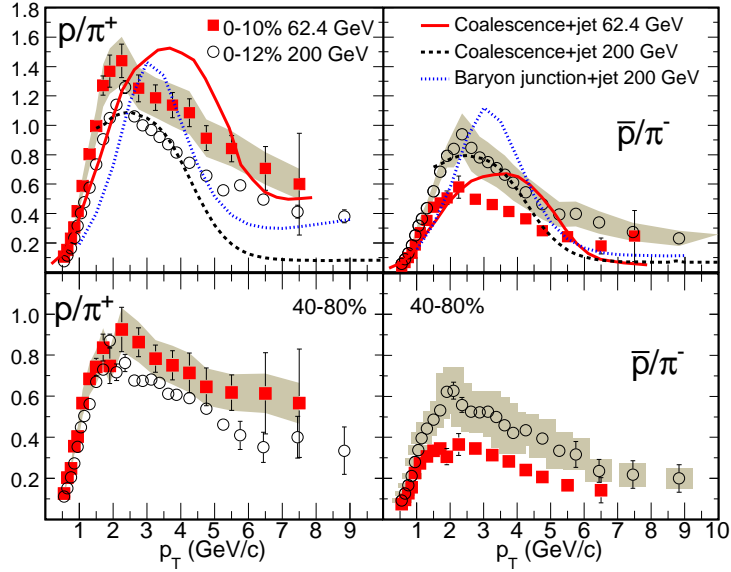}
\caption{$p/\pi^{+}$ and $\bar{p}$/$\pi^{-}$ ratios
versus $p_{\mathrm T}$
at $\sqrt{s_{\mathrm {NN}}}$ = 62.4 and 200 GeV for central (upper panels)
and peripheral (lower panels) collisions. 
For clarity of presentation, the systematic errors (shaded bands) are 
shown for only one of the beam energy for a given ratio. They are of similar 
magnitude at the other beam energy. 
The curves are model results~\cite{vitev_62,baryonjunc,fries} 
and are discussed in the text.
}
\label{fig4}
\end{center}
\end{figure}

Figure~\ref{fig4} shows the $p/\pi^{+}$ and $\bar{p}$/$\pi^{-}$ ratios
versus $p_{\mathrm T}$ for Au+Au 0-10\% and 0-12\% (upper panels),
and  40-80\% (lower panels) centralities
at $\sqrt{s_{\mathrm {NN}}}$ = 62.4 GeV and 200 GeV,
together with theoretical predictions to be discussed.

    The fact that for central collisions the $p/\pi^{+}$ and $\bar{p}$/$\pi^{-}$ ratios are close to unity
    in the intermediate $p_{\mathrm T}$ region at
    200 GeV has been attributed to either
    quark coalescence~\cite{reco,vitev_62} or novel baryon transport
    dynamics based on topological gluon field configurations~\cite{baryonjunc}.
    The quark coalescence models predict a specific
    energy dependence for  $p/\pi^{+}$,  being higher
    at $\sqrt{s_{\mathrm {NN}}}$ = 62.4 GeV than at
    200 GeV in the intermediate $p_{\mathrm T}$ region; the
    energy dependence is reversed for $\bar{p}$/$\pi^{-}$~\cite{vitev_62}.
    On the other hand, the  baryon junction model predicts a decrease
    in the ratio at intermediate $p_{\mathrm T}$ with decreasing
    collision centrality at a given $\sqrt{s_{\mathrm {NN}}}$~\cite{baryonjunc}.

    As Fig.~\ref{fig4} shows, at a given $p_{\mathrm T}$
    the $p/\pi^{+}$ ratio
    for Au+Au at 62.4 GeV is larger than
    the value at 200 GeV
    in the intermediate $p_{\mathrm T}$ range,
    whereas  for $\bar{p}$/$\pi^{-}$ the reverse occurs. This
    specific energy dependence of the baryon-to-meson ratio
    as a function of $p_{\mathrm T}$ is consistent with the general
    expectation from quark coalescence models~\cite{vitev_62}.
    Our results also show that the baryon-to-meson ratios, here 
    $p/\pi^{+}$, for the region 1.5 $<$ $p_{\mathrm T}$ $<$ 6 GeV/$c$
    are higher than in $p$+$p$ and $d$+Au~\cite{ppdau}.
    This enhancement increases with centrality for both beam energies.

The ratios for the 0-10\% and 0-12\% centrality data
(upper panels of Fig.~\ref{fig4}) are compared to
predictions from models based on quark coalescence and a jet
fragmentation mechanism for particle production at 62.4
GeV~\cite{vitev_62} and 200 GeV~\cite{fries}, and baryon junction
and jet fragmentation at 200 GeV~\cite{baryonjunc}. 
For the intermediate $p_{\mathrm T}$ region there is a lack of quantitative
agreement between  model results and data. The recombination
models predict a shift in the peak position of the ratios to higher 
$p_{\mathrm T}$ at the 62.4 GeV, which is not
observed. The $\bar{p}/\pi^{-}$ ratios for the two energies do not cross-over 
as predicted by the models.
The baryon junction model predictions are not in quantitative 
agreement with our  200 GeV data.
%%%%%%%%%%%%%% Fig. 4 %%%%%%%%%%%%%%%%%%%%%%%%%%%%%
%%%%%%%%%% End of Fig.4 %%%%%%%%%%%%%%%%%%%%%%%%%%%%%

    At higher $p_{\mathrm T}$ the $p/\pi^{+}$ and $\bar{p}/\pi^{-}$ ratios
    are nearly independent of centrality
    at both 62.4 and 200 GeV.
    This observation, taken together with a constant $R_{\mathrm {CP}}$
    beyond $p_{\mathrm T}$~$>$~6 GeV/$c$, may reflect 
    the dominance of particle production from the fragmentation
    mechanism. Also, at high $p_{\mathrm T}$ ($p_{\mathrm T}$~$>$~6~GeV/$c$)
    we observed a similar $\bar{p}/\pi^{-}$ ratio in central Au+Au 
    and $d$+Au at 200 GeV~\cite{star_pid200,ppdau} and a similar $R_{\mathrm CP}$ for $p$+$\bar{p}$ 
    and $\pi^{+}+\pi^{-}$ for Au+Au (see previous section). These observations
    appear to be inconsistent with the naive expectations from the color charge dependence
    of the parton energy loss~\cite{xnwang_nonabelian,th_prob}. The difference in quark and gluon
    energy loss would have led to a lower $\bar{p}/\pi^{-}$ ratio for Au+Au
    at high $p_{\mathrm T}$ than that for $d$+Au collisions and a lower $R_{\mathrm CP}$ 
    for $p$+$\bar{p}$ compared to $\pi^{+}+\pi^{-}$. Recent theoretical calculations 
    suggest that a much larger net quark to gluon jet conversion rate in the QGP
    medium is needed than given by the lowest order QCD calculations to
    explain the high $p_{\mathrm T}$ particle ratios~\cite{ko}.
   
%%%%%%%%%%%%%% Fig. 5 %%%%%%%%%%%%%%%%%%%%%%%%%%%%%
\begin{figure}
\begin{center}
\includegraphics[scale=0.4]{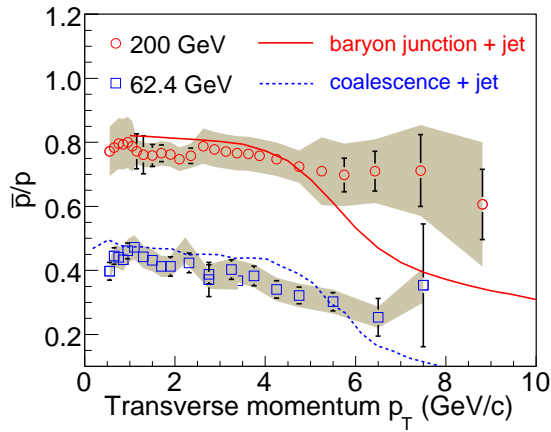}
\caption{
The $\bar{p}/p$ ratios versus $p_{\mathrm T}$ at $\sqrt{s_{\mathrm {NN}}}$ = 62.4 (0-10\%)
and 200 GeV (0-12\%). The errors shown are
statistical, and the shaded bands reflect the systematic errors.
Model predictions are shown as solid and dashed curves for 200 GeV~\cite{baryonjunc}
and 62.4 GeV~\cite{vitev_62} central Au+Au, respectively.
}
\label{fig5}
\end{center}
\end{figure}
%%%%%%%%%% End of Fig.5 %%%%%%%%%%%%%%%%%%%%%%%%%%%%%

   As one can see in Fig.~\ref{fig4}, the jet fragmentation prediction is reasonable at
   high $p_{\mathrm T}$ for the $p/\pi^{+}$ ratios at 62.4 GeV. 
   However these calculations predict a much lower value for the ratio at 200 GeV. 
   The failure of these model calculations at high $p_{\mathrm T}$ is further 
   noticeable when we compare them to the measured $\bar{p}/p$ ratios at both
   energies. Figure~\ref{fig5} shows the $\bar{p}/p$ ratios versus $p_{\mathrm T}$ at 62.4 and 
   200 GeV. The data are compared to a model result in which baryons and anti-baryons are produced 
   through baryon junctions  and jet fragmentation at 200 GeV~\cite{baryonjunc} 
   and through coalescence and jet fragmentation processes
   at 62.4 GeV~\cite{vitev_62}. Both the models overpredict the data at lower $p_{\mathrm T}$ 
   ($p_{\mathrm T}$~$<$~5~GeV/$c$). For $p_{\mathrm T}$~$>$~6 GeV/$c$, where fragmentation is 
   the dominant mechanism of particle production in the models, they underpredict the measured $\bar{p}/p$ 
   ratios at the two beam energies. The model calculations do not use the recent fragmentation
   functions for $p$+$\bar{p}$ as supported by the RHIC data from $p$+$p$ and $d$+Au collisions
   at 200 GeV~\cite{ppdau}.

\begin{figure}
\begin{center}
\includegraphics[scale=0.4]{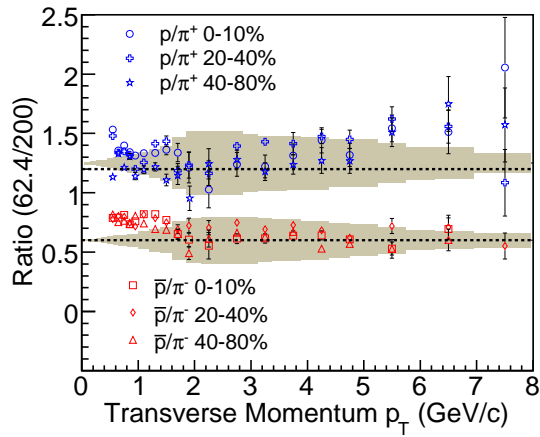}
\caption{
The ratios $p/\pi^{+}$ at $\sqrt{s_{NN}}$ = 62.4 GeV and 200 GeV
and $\bar{p}/\pi^{-}$ at $\sqrt{s_{NN}}$ = 62.4 GeV and 200 GeV
as functions of $p_{\mathrm T}$ for various collision centrality classes.
The error bars are statistical and shaded bands are the systematic errors.
}
\label{fig6}
\end{center}
\end{figure}

To further investigate the energy dependence of baryon-to-meson ratios, we 
present the ratio of $\bar{p}/\pi^{-}$ between 62.4 GeV and 200 GeV and 
the ratio of $p/\pi^{+}$ between 62.4 GeV and 200 GeV.
Figure~\ref{fig6} shows that this double ratio of $\bar{p}/\pi^{-}$ is
independent of $p_{\mathrm T}$ with a value around 0.6 for $p_{T}>2$ GeV/$c$, while the double ratio of
$p/\pi^{+}$ is around 1.2 for $p_{T}\simeq2 - 5$ GeV/$c$ and increases with
$p_T$, possibly due to different valence quark contributions at the two energies.
%%%%%%%%%%%%%% Fig. 6 %%%%%%%%%%%%%%%%%%%%%%%%%%%%%
%%%%%%%%%% End of Fig.6 %%%%%%%%%%%%%%%%%%%%%%%%%%%%%

Baryon and meson production at high $p_T$ and the relative contributions from quark and
gluon jets have been discussed in Refs.~\cite{star_pid200,ppdau}.
The new observations presented in this paper necessitate further 
understanding of the role of gluons, quarks and their energy loss mechanisms. 
Gluon jets tend to produce more baryons than quark jets, whereas quark jets 
contributes substantially to pion production. This feature is supported by a 
much lower  $\bar{p}/\pi^{-}$ ratio at high $p_{\mathrm T}$ compared to the
 $p/\pi^{+}$ ratio for low-energy $p$+$p$ and $p$+A collisions~\cite{ppdau}. 
The double ratios in Fig.~\ref{fig6} are consistent with this picture as the
$\bar{p}/\pi^{-}$ ratio is lower at 62 GeV than at 200 GeV. 
Due to their larger coupling gluons should lose more energy in the dense medium formed in heavy-ion 
collisions than quarks~\cite{xnwang_nonabelian,th_prob}.
This would lead to a lower $\bar{p}/\pi^{-}$ ratio for central
Au+Au relative to peripheral Au+Au at both beam energies. This is not observed 
for the data reported here.
A larger value of the $p/\pi^{+}$ ratio at 62 GeV than at 200 GeV 
is observed. This may be due to greater valence quark contribution at the
lower beam energy. However, the $p/\pi^{+}$ double ratio shows no centrality dependence. 
This is not expected if valence quarks contribute significantly more at lower energy
and lose energy in the dense medium formed for central Au+Au.

At intermediate $p_{\mathrm T}$, the features of the double ratios are 
not expected from the coalescence model;  as seen in
Fig.~\ref{fig4} the quark coalescence models will lead to more prominent
baryon enhancement at 62 GeV than at 200 GeV. It is, however,
surprising that the scaling is independent of centrality and extends
to high $p_{\mathrm T}$ when baryons are more enhanced at
intermediate $p_{\mathrm T}$.

\section{Summary}

We have presented a study of the energy dependence of 
$\pi^{\pm}$, $p$ and $\bar{p}$ production for Au+Au at
$\sqrt{s_{\mathrm {NN}}}$ = 62.4 and 200 GeV. The $p_{\mathrm T}$
spectra are measured around midrapidity ($\mid$$y$$\mid$~$<$~0.5)
over the range  0.2~$<$~$p_{\mathrm T}$~$<$~12 GeV/$c$.
These measurements provide new experimental data for investigating the production
of quarks, gluons and their interactions with the medium formed in heavy-ion collisions
and the interplay between coalescence of thermal partons and jet fragmentation.

The $p_{\mathrm T}$ dependence of $R_{\mathrm CP}$ for charged pions and 
for protons and anti-protons is different at both energies. However, at 
higher $p_{\mathrm T}$ the values of
$R_{\mathrm {CP}}$ for baryons and mesons at both energies are
similar. The comparison of $R_{\mathrm AA}$ versus $p_{\mathrm T}$ to model predictions 
provides important information on quantities like initial gluon density 
and lifetime of dense matter.

The $p$/$\pi^{+}$ ratios for Au+Au  at $\sqrt{s_{\mathrm
{NN}}}$ = 62.4 GeV are higher than the corresponding values at
$\sqrt{s_{\mathrm {NN}}}$ = 200 GeV in the intermediate  $p_{\mathrm
T}$ range, but the $\bar{p}$/$\pi^{-}$ ratios are smaller. There
is serious quantitative disagreement between data and the available theoretical models.
We observe a scaling of the $\bar{p}$/$\pi^{-}$ ratios between  corresponding centralities for
the two beam energies at $p_T>2$ GeV/c despite the strong centrality
and $p_T$ dependence of these ratios.

We thank J. Alam, V. Greco, C.M. Ko, I. Vitev and X.-N. Wang 
for providing the theoretical results for comparison with the data.
We thank the RHIC Operations Group and RCF at BNL, and the
NERSC Center at LBNL for their support. This work was supported
in part by the Offices of NP and HEP within the U.S. DOE Office 
of Science; the U.S. NSF; the BMBF of Germany; CNRS/IN2P3, RA, RPL, and
EMN of France; EPSRC of the United Kingdom; FAPESP of Brazil;
the Russian Ministry of Science and Technology; the Ministry of
Education and the NNSFC of China; IRP and GA of the Czech Republic,
FOM of the Netherlands, DAE, DST, and CSIR of the Government
of India; Swiss NSF; the Polish State Committee for Scientific 
Research; SRDA of Slovakia, and the Korea Sci. \& Eng. Foundation.


\begin{thebibliography}{99}
\bibitem{star_scaling} Earlier it was reported that neutral pions show  $x_{\mathrm
T}$-scaling  in the region 0.03 $<$ $x_{\mathrm T}$ $<$ 0.06 for
$\sqrt{s_{\mathrm {NN}}}$ = 130 GeV and 200 GeV with an effective
scaling  power of 6.41 $\pm$ 0.49~\cite{phenix_scaling}. 
We did a similar study for the charged pions at $\sqrt{s_{\mathrm {NN}}}$ = 62.4 GeV
and 200 GeV and found  $x_{\mathrm T}$-scaling  in a different region 0.06
$<$ $x_{\mathrm T}$ $<$ 0.11 with a higher effective scaling power of  7.4
$\pm$ 0.2 for central Au+Au.
\end{thebibliography}
\end{document}